\documentclass[reqno]{amsart}

\makeatletter
\@addtoreset{equation}{section}

\makeatother

\newcommand{\Cm}{\mathbb{C}}

\newcommand{\be}{\begin{equation}}
\newcommand{\ee}{\end{equation}}
\newcommand{\va}{\varphi}

\newcommand{\pp}{\partial}

\usepackage{amsmath}
\usepackage[foot]{amsaddr}
\usepackage{amsthm,amssymb,mathrsfs}
\usepackage{graphicx}
\usepackage{color}

\newtheorem{thm}{Theorem}[section]

\theoremstyle{remark}\newtheorem{rmk}[thm]{Remark}

\title[]{Linear transport in porous media}

\author{Kenji Amagai$^1$}
\author{Yuko Hatano$^1$}
\address{${^1}$Graduate School of Systems and Information Engineering, 
University of Tsukuba, Tsukuba 305-7361, Japan}

\author{Manabu Machida$^2$}
\address{${^2}$Institute for Medical Photonics Research, 
Hamamatsu University School of Medicine, 
Hamamatsu 431-3192, Japan}
\email{machida@hama-med.ac.jp}

\date{\today}

\begin{document}

\begin{abstract}
The linear transport theory is developed to describe the time dependence of the number density of tracer particles in porous media. The advection is taken into account. The transport equation is numerically solved by the analytical discrete ordinates method. For the inverse Laplace transform, the double-exponential formula is employed.
\end{abstract}

\maketitle

\section{Introduction}

The use of the transport equation for the flow in porous media was proposed by Williams \cite{Williams_1992a,Williams_1992b,Williams_1993a,Williams_1993b}. Recently it was experimentally shown that the concentration of tracer particles in column experiments obeys the transport equation \cite{Amagai-etal20}. In \cite{Amagai-etal20}, the advection $u>0$ was taken into account. Then the spatial derivative term in the transport equation is given by $(u+v_0\mu)\pp/\pp x$ instead of $\mu\pp/\pp x$, where $v_0>0$ is the inherent particle speed and $\mu\in[-1,1]$ is the cosine of the polar angle. The transport equation with such a spatial derivative term has been explored in the context of the evaporation of rarefied gas \cite{Loyalka-etal81,Scherer-Barichello09,Siewert-Thomas81,Siewert-Thomas82}.

In column experiments \cite{Cortis-etal04}, a column tube is filled with a solute such as sands or glass beads, and water is poured from the top end of the column with a constant pressure. Then tracer particles are injected to the water. They enter the column from the top and eventually exit from the bottom of the column. In \cite{Amagai-etal20}, the concentration of tracer particles was computed making use of the method of analytical discrete ordinates (ADO). Both the short-time growing behavior and long-time decay behavior of the breakthrough curve (the time dependence of the concentration) were well reproduced by the transport equation \cite{Amagai-etal20}.

In \cite{Amagai-etal20}, the solution to the transport equation for a semi-infinite medium was compared to the experimental data. In this paper, we give a formulation for the transport equation in the slab geometry taking into account the length of the column. We make use of the double-exponential formula for the inverse Laplace transform.

The rest of the paper is organized as follows. In Sec.~\ref{sec:Method}, we formulate our transport problem in the slab geometry. In Sec.~\ref{sec:ADO}, a numerical scheme is developed using ADO. In Sec.~\ref{sec:invLaplace}, our numerical scheme of the inverse Laplace transform is described. Finally, concluding remarks are given in Sec.~\ref{concl}.

\section{The transport equation}
\label{sec:Method}

Let us consider the one-dimensional linear Boltzmann equation. The velocity $v$ is given by $v=u+v_0\mu$. Let $\sigma_a>0$ and $\sigma_s>0$ be the absorption and scattering coefficients, respectively. Let $L$ be the length of the column. We write our transport equation as follows.
\be
\left\{\begin{aligned}
&
P\psi(x,\mu,t)=
\frac{\sigma_s}{2}\int^{1}_{-1}\psi(x,\mu,t)\,d\mu,
\quad 0<x<L,\quad-1\le\mu\le1,\quad t>0,
\\
&
\psi(x,\mu,0)=0,\quad 0<x<L,\quad-1\le\mu\le1,
\\
&
\psi(0,\mu,t)=n_0\delta(\mu-1),\quad-\eta<\mu\le1,\quad t>0,
\\
&
\psi(L,\mu,t)=0,\quad-1\le\mu<-\eta,\quad t>0,
\end{aligned}\right.
\label{rte0}
\ee
where
\[
P\psi(x,\mu,t)=
\left[\frac{\pp}{\pp t}+(u+v_0\mu)\frac{\pp}{\pp x}+\sigma_a+\sigma_s\right]\psi(x,\mu,t).
\]
Here, $\delta(\mu-1)$ is the Dirac delta function, $n_0$ is the initial particle number density, and
\[
\eta=\frac{u}{v_0}.
\]
We call $\psi(x,\mu,t)$ the angular number density. The particle number density $n(t)$ at $x=L$ is given by
\[
n(t)=\int_{-\eta}^1\psi(L,\mu,t)\,d\mu.
\]

\section{The Laplace transform}
\label{sec:ADO}

Let us introduce the Laplace transform
\[
\hat{\psi}(x,\mu,p)=\int_0^{\infty}e^{-pt}\psi(x,\mu,t)\,dt,
\quad p\in\Cm,
\]
and new variables
\[
\mu_t=\frac{\sigma_a+\sigma_s+p}{v_0}\in\Cm,\quad
\mu_s=\frac{\sigma_s}{v_0}>0.
\]
Then we have
\[
\left\{\begin{aligned}
&
(\eta+\mu)\frac{\pp}{\pp x}\hat{\psi}(x,\mu,p)+\mu_t\hat{\psi}(x,\mu,p)=\frac{\mu_s}{2}\int_{-1}^1\hat{\psi}(x,\mu,p)\,d\mu,
\quad0<x<L,\quad-1\le\mu\le1,
\\
&
\hat{\psi}(0,\mu,p)=\frac{n_0}{p}\delta(\mu-1),\quad-\eta<\mu\le1,
\\
&
\hat{\psi}(L,\mu,p)=0,\quad-1\le\mu<-\eta.
\end{aligned}\right.
\]
We note that the coefficient $\mu_t$ is a complex number and the boundary conditions are specified by intervals $(-\eta,1]$ and $[-1,-\eta)$. We will carry out the numerical computation for the above-mentioned equation using the analytical discrete ordinates method (ADO) \cite{Barichello_2000,Barichello_2001,Siewert_1999}.

\begin{rmk}
By changing $\mu+\eta\to\tilde{\mu}$ and defining $\hat{f}(x,\tilde{\mu},p)=\hat{\psi}(x,\tilde{\mu}-\eta)$, we can reformulate the equation as
\[
\left\{\begin{aligned}
&
\tilde{\mu}\frac{\pp}{\pp x}\hat{f}(x,\tilde{\mu},p)+\mu_t\hat{f}(x,\tilde{\mu},p)=\frac{\mu_s}{2}\int_{\eta-1}^{\eta+1}\hat{f}(x,\tilde{\mu},p)\,d\tilde{\mu},
\quad0<x<L,\quad\eta-1\le\tilde{\mu}\le\eta+1,
\\
&
\hat{f}(0,\tilde{\mu},p)=\frac{n_0}{p}\delta(\tilde{\mu}-\eta-1),
\quad0<\tilde{\mu}\le\eta+1,
\\
&
\hat{f}(L,\tilde{\mu},p)=0,
\quad\eta-1\le\tilde{\mu}<0,
\end{aligned}\right.
\]
Such transformation is particularly useful for the evaporation problem, in which the integral on the right-hand side of the transport equation is taken from $-\infty$ to $\infty$ \cite{Scherer-Barichello09}.
\end{rmk}

Let us write
\[
\hat{\psi}(x,\mu,p)=\hat{\psi}_b(x,\mu,p)+\hat{\psi}_s(x,\mu,p).
\]
The ballistic term $\hat{\psi}_b$ satisfies
\[
\left\{\begin{aligned}
&
(\eta+\mu)\frac{\pp}{\pp x}\hat{\psi}_b(x,\mu,p)+\mu_t\hat{\psi}_b(x,\mu,p)=0,
\quad0<x<L,\quad-1\le\mu\le1,
\\
&
\hat{\psi}_b(0,\mu,p)=\frac{n_0}{p}\delta(\mu-1),\quad-\eta<\mu\le1,
\\
&
\hat{\psi}_b(L,\mu,p)=0,\quad-1\le\mu<-\eta,
\end{aligned}\right.
\]
and the scattering term $\hat{\psi}_s$ obeys
\[
\left\{\begin{aligned}
&
\left[(\eta+\mu)\frac{\pp}{\pp x}+\mu_t\right]\hat{\psi}_s=\frac{\mu_s}{2}\int_{-1}^1\hat{\psi}_s(x,\mu,p)\,d\mu+q,
\quad0<x<L,\quad-1\le\mu\le1,
\\
&
\hat{\psi}_s(0,\mu,p)=0,\quad-\eta<\mu\le1,
\\
&
\hat{\psi}_s(L,\mu,p)=0,\quad-1\le\mu<-\eta,
\end{aligned}\right.
\]
where
\[
q(x,\mu,p)=\frac{\mu_s}{2}\int^{1}_{-1}\hat{\psi}_b(x,\mu,p)\,d\mu=
\frac{n_0\mu_s}{2p}e^{-x\mu_t/(\eta+1)}.
\]
We note that
\[
\hat{\psi}_b(x,\mu,p)=
\frac{n_0}{p}e^{-x\mu_t/(\eta+\mu)}\delta(\mu-1).
\]

Let us express $\hat{\psi}_s(x,\mu,p)=\hat{\psi}_s(x,\mu)$ and $q(x,\mu,p)=q(x,\mu)$ when there is no confusion. For the computation of $\hat{\psi}_s$, we discretize the integral by the Gauss-Legendre quadrature and obtain
\[
\left[(\eta+\mu_i)\frac{\pp}{\pp x}+\mu_t\right]\hat{\psi}_s(x,\mu_i)=
\frac{\mu_s}{2}\sum_{j=1}^Nw_j
\left[\hat{\psi}_s(x,\mu_j)+\hat{\psi}_s(x,-\mu_j)\right]+q(x,\mu_i),
\]
where $\mu_i,w_i$ ($i=1,2,\dots,2N$) are abscissas and weights, respectively. We have $0<\mu_1<\cdots<\mu_N<1$ and $\mu_{N+i}=-\mu_i$ ($i=1,\dots,N$). Furthermore, we introduce $N_{\eta}$ as the largest integer such that $-\eta<\mu_{N_{\eta}}$.

\begin{rmk}
It is possible to assign different abscissas and weights for two intervals $[-1,-\eta)$ and $(-\eta,1]$. Since we assume $\eta$ is small, we use one set of abscissas and weights for the interval $[-1,1]$ as described above.
\end{rmk}

The scattering part $\hat{\psi}_s$ is obtained as
\[
\hat{\psi}_s(x,\mu_i)=
\sum^{2N}_{j=1}\int_0^LG(x,\mu_i;x',\mu_j)q(x',\mu_j)\,dx',
\]
where the Green's function defined for each $p$ satisfies
\[
\left\{\begin{aligned}
&
\left[(\eta+\mu_i)\frac{\pp}{\pp x}+\mu_t\right]G(x,\mu_i;x',\mu_j)=
\frac{\mu_s}{2}\sum^{2N}_{k=1}w_kG(x,\mu_k;x',\mu_j)+\delta(x-x')\delta_{ij},
\\
&
G(0,\mu_i;x',\mu_j)=0,\quad\mu_i>\mu_{N_{\eta}},
\\
&
G(L,\mu_i;x',\mu_j)=0,\quad\mu_i<\mu_{N_{\eta}},
\end{aligned}\right.
\]
where $\delta_{ij}$ is the Kronecker delta.

Let us consider the following homogeneous equation.
\[
\left((\eta+\mu_i)\frac{\pp}{\pp x}+\mu_t\right)\hat{\psi}(x,\mu_i)=\frac{\mu_s}{2}\sum_{j=1}^{2N}w_j\hat{\psi}(x,\mu_j).
\]
We note that $\hat{\psi}$ depends on $p$ through $\mu_t$. With separation of variables, we can write $\hat{\psi}$ as
\[
\hat{\psi}(x,\mu_i)=\phi(\nu,\mu_i)e^{-x/\nu},
\]
where $\nu$ is the separation constant. The function $\phi(\nu,\mu_i)$ satisfies the normalization condition,
\[
\sum_{i=1}^{2N}w_i\phi(\nu,\mu_i)=
\sum_{i=1}^Nw_i\left(\phi(\nu,\mu_i)+\phi(\nu,-\mu_i)\right)=1.
\]
We obtain
\[
\phi(\nu,\mu_i)=\frac{\mu_s\nu}{2}\frac{1}{\mu_t\nu-\mu_i-\eta},
\]
assuming $\nu\neq(\mu_i+\eta)/\mu_t$. If $\eta=0$ and $p$ is real, we can prove $\mu\neq\mu_i/\mu_t$ \cite{Siewert_1999}. The following orthogonality relation holds.
\[
\sum_{i=1}^{2N}w_i(\mu_i+\eta)\phi(\nu,\mu_i)\phi(\nu',\mu_i)=
\mathcal{N}(\nu)\delta_{\nu\nu'},
\]
where
\[
\mathcal{N}(\nu)=
\sum_{i=1}^{2N}w_i(\mu_i+\eta)\phi(\nu,\mu_i)^2.
\]
We can find $2N$ eigenvalues $\nu=\nu_n$ ($n=1,2,\dots,2N$). Moreover there are $N_{\eta}$ eigenvalues with positive real parts. See \cite{Amagai-etal20} for the computation of eigenvalues. Moreover, the free-space Green's function $G_0$ is obtained as
\[
G_0(x,\mu_i;x',\mu_j)=
\pm\sum_{\pm\Re{\nu_n}>0}\frac{w_j}{\mathcal{N}(\nu_n)}
\phi(\nu_n,\mu_j)\phi(\nu_n,\mu_i)e^{-(x-x')/\nu_n},
\]
where upper signs are chosen for $x>x'$ and lower signs are used for $x<x'$.

Hence we can write
\[
\begin{aligned}
G(x,\mu_i;x',\mu_j)
&=
G_0(x,\mu_i;x',\mu_j)
\nonumber \\
&+
\sum_{\Re{\nu_n}>0}B_1(\nu_n)\phi(\nu_n,\mu_i)e^{-x/\nu_n}+
\sum_{\Re{\nu_n}<0}B_2(\nu_n)\phi(\nu_n,\mu_i)e^{-x/\nu_n},
\end{aligned}
\]
where coefficients $B_1(\nu_n),B_2(\nu_n)$ are determined from boundary conditions. We obtain
\begin{align}
\sum_{\Re{\nu_n}>0}B_1(\nu_n)\phi(\nu_n,\mu_{i_1})+\sum_{\Re{\nu_n}<0}B_2(\nu_n)\phi(\nu_n,\mu_{i_1})
&=
y_1(\mu_{i_1}),
\label{y1eq}
\\
\sum_{\Re{\nu_n}>0}B_1(\nu_n)\phi(\nu_n,\mu_{i_2})e^{-L/\nu_n}+\sum_{\Re{\nu_n}<0}B_2(\nu_n)\phi(\nu_n,\mu_{i_2})e^{-L/\nu_n}
&=
y_2(\mu_{i_2}),
\label{y2eq}
\end{align}
where $1\le i_1\le N_{\eta}$, $N_{\eta}<i_2\le 2N$,
\[
\begin{aligned}
y_1(\mu_{i_1})&=
\sum_{\Re{\nu_n}<0}\frac{w_j}{\mathcal{N}(\nu_n)}
\phi(\nu_n,\mu_j)\phi(\nu_n,\mu_{i_1})e^{x'/\nu_n},
\\
y_2(\mu_{i_2})&=
-\sum_{\Re{\nu_n}>0}\frac{w_j}{\mathcal{N}(\nu_n)}
\phi(\nu_n,\mu_j)\phi(\nu_n,\mu_{i_2})e^{-(L-x')/\nu_n}.
\end{aligned}
\]
Let us multiply (\ref{y1eq}) and (\ref{y2eq}) by $\exp(-x'\mu_t/(\eta+1))$, integrate both sides of these equations over $x'$, and take the sum with respect to $j$. We obtain
\begin{align}
\sum_{\Re{\nu_n}>0}E_1(\nu_n)\phi(\nu_n,\mu_{i_1})+\sum_{\Re{\nu_n}<0}E_2(\nu_n)\phi(\nu_n,\mu_{i_1})
&=
z_1(\mu_{i_1}),
\label{z1eq}
\\
\sum_{\Re{\nu_n}>0}E_1(\nu_n)\phi(\nu_n,\mu_{i_1})e^{-L/\nu_n}+\sum_{\Re{\nu_n}<0}E_2(\nu_n)\phi(\nu_n,\mu_{i_2})e^{-L/\nu_n}
&=
z_2(\mu_{i_2}),
\label{z2eq}
\end{align}
where
\[
\begin{aligned}
z_1(\mu_{i_1})&=
\sum_{\Re{\nu_n}<0}\frac{(\eta+1)\nu_n}{\mathcal{N}(\nu_n)(\eta+1-\nu_n\mu_t)}
\left[e^{L/\nu_n-L\mu_t/(\eta+1)}-1\right]\phi(\nu_n,\mu_{i_1}),
\\
z_2(\mu_{i_2})&=
\sum_{\Re{\nu_n}>0}\frac{(\eta+1)\nu_n}{\mathcal{N}(\nu_n)(\eta+1-\nu_n\mu_t)}
\left[e^{L/\nu_n-L\mu_t/(\eta+1)}-1\right]e^{-L/\nu_n}\phi(\nu_n,\mu_{i_2}),
\end{aligned}
\]
and
\[
E_k(\nu_n)=
\sum_{j=1}^{2N}\int_0^LB_k(\nu_n)e^{-\mu_tx'/(\eta+1)}\,dx'
\quad(k=1,2).
\]
Thus $E_1(\nu_n),E_2(\nu_n)$ are obtained from (\ref{z1eq}) and (\ref{z2eq}).

Hence we obtain
\[
\begin{aligned}
\hat{\psi}_s(x,\mu_i)
&=
\frac{n_0\mu_s}{2p}\sum_{j=1}^{2N}\int_0^LG(x,\mu_i;x',\mu_j)e^{-x'\mu_t/(\eta+1)}\,dx'
\\
&=
\frac{n_0\mu_s}{2p}\sum_{\Re{\nu_n}>0}
\frac{(\eta+1)\nu_n}{\mathcal{N}(\nu_n)(\eta+1-\nu_n\mu_t)}
\left(e^{-\mu_tx/(\eta+1)}-e^{-x/\nu_n}\right)\phi(\nu_n,\mu_i)
\\
&+
\frac{n_0\mu_s}{2p}\sum_{\Re{\nu_n}<0}
\frac{(\eta+1)\nu_n}{\mathcal{N}(\nu_n)(\eta+1-\nu_n\mu_t)}
e^{-x\mu_t/(\eta+1)}\left(1-e^{(L-x)/\nu_n-(L-x)\mu_t/(\eta+1)}\right)
\phi(\nu_n,\mu_i)
\\
&+
\frac{n_0\mu_s}{2p}\left[\sum_{\Re{\nu_n}>0}E_1(\nu_n)\phi(\nu_n,\mu_i)e^{-x/\nu_n}+\sum_{\Re{\nu_n}<0}E_2(\nu_n)\phi(\nu_n,\mu_i)e^{-x/\nu_n}\right].
\end{aligned}
\]
The Laplace transform of $n(t)$ is obtained as
\[
\begin{aligned}
\hat{n}(p)&=
\int_{-\eta}^1\hat{\psi}(L,\mu,p)\,d\mu
\\
&=
\int_{-\eta}^1\hat{\psi}_b(L,\mu,p)\,d\mu+
\int_{-\eta}^1\hat{\psi}_s(L,\mu,p)\,d\mu.
\end{aligned}
\]
Therefore,
\[
\begin{aligned}
\hat{n}(p)&=
\frac{n_0}{p}e^{-L\mu_t/(\eta+1)}
\\
&+
\frac{n_0\mu_s}{2p}\sum_{\Re{\nu_n}>0}
\frac{(\eta+1)\nu_n}{\mathcal{N}(\nu_n)(\eta+1-\nu_n\mu_t)}
\left(e^{-L\mu_t/(\eta+1)}-e^{-L/\nu_n}\right)\va(\nu_n)
\\
&+
\frac{n_0\mu_s}{2p}\left[\sum_{\Re{\nu_n}>0}E_1(\nu_n)e^{-L/\nu_n}\va(\nu_n)+\sum_{\Re{\nu_n}<0}E_2(\nu_n)e^{-L/\nu_n}\va(\nu_n)\right],
\end{aligned}
\]
where
\[
\va(\nu)=
\sum_{i=1}^{N_{\eta}}w_i\phi(\nu,\mu_i)=
\frac{\mu_s\nu}{2}\sum_{i=1}^{N_{\eta}}\frac{w_i}{\mu_t\nu-\mu_i-\eta}.
\]
By the inverse Laplace transform, we have
\be
\begin{aligned}
n(t)
&=
\frac{n_0}{2\pi i}\int_{\gamma-i\infty}^{\gamma+i\infty}\frac{e^{pt}}{p}
\Biggl\{
e^{-L\mu_t/(\eta+1)}
\\
&+
\frac{\mu_s}{2}\sum_{\Re{\nu_n}>0}
\frac{(\eta+1)\nu_n}{\mathcal{N}(\nu_n)(\eta+1-\nu_n\mu_t)}
\left(e^{-L\mu_t/(\eta+1)}-e^{-L/\nu_n}\right)\va(\nu_n)
\\
&+
\frac{\mu_s}{2}\left[\sum_{\Re{\nu_n}>0}E_1(\nu_n)e^{-L/\nu_n}\va(\nu_n)+\sum_{\Re{\nu_n}<0}E_2(\nu_n)e^{-L/\nu_n}\va(\nu_n)\right]
\Biggr\}\,dp,
\end{aligned}
\label{ntsol}
\ee
where $\gamma$ is taken to be greater than the largest real part of any singularity.

\section{The inverse Laplace transform}
\label{sec:invLaplace}

Let us numerically evaluate the Bromwich integral in the inverse Laplace transform (\ref{ntsol}). Although the trapezoidal rule was used in \cite{Amagai-etal20}, here we employed the double-exponential formula \cite{Ooura-Mori91,Ooura-Mori99}.

We note that $\Im\hat{n}(\gamma+i\omega/t)=-\Re\hat{n}(\gamma+i(\omega-\frac{\pi}{2})/t)$ because
\[
\hat{n}\left(\gamma+i\frac{\omega-\frac{\pi}{2}}{t}\right)
=\int_0^{\infty}e^{-\gamma t-i(\omega-\frac{\pi}{2})}n(t)\,dt
=i\hat{n}\left(\gamma+i\frac{\omega}{t}\right).
\]
For $t>0$, we have
\be
\begin{aligned}
n(t)
&=
\frac{e^{\gamma t}}{2\pi t}\int_{-\infty}^{\infty}e^{i\omega}\hat{n}\left(\gamma+i\frac{\omega}{t}\right)\,d\omega=
\frac{e^{\gamma t}}{\pi t}\int_{-\infty}^{\infty}\cos\omega\,\Re\hat{n}\left(\gamma+i\frac{\omega}{t}\right)\,d\omega
\\
&=
\frac{2e^{\gamma t}}{\pi t}\int_0^{\infty}\cos\omega\,\Re\hat{n}\left(\gamma+\frac{i\omega}{t}\right)\,d\omega.
\end{aligned}
\label{ntcosw}
\ee
Let us introduce $\phi(\tau)=\tau/[1-e^{-6\sinh{\tau}}]$. Then the above integral can be written as
\[
n(t)=
\frac{2e^{\gamma t}}{\pi t}\int_{-\infty}^{\infty}\cos(M\phi(\tau))\,\Re\hat{n}\left(\gamma+\frac{i}{t}M\phi(\tau)\right)M\frac{d}{d\tau}\phi(\tau)\,d\tau,
\]
where $\omega=M\phi(\tau)$, $M>0$. Let us define $h=\pi/M$. By the trapezoidal rule, we arrive at
\be
n(t)\approx
\frac{2e^{\gamma t}}{t}\sum_{k=-k_{\rm max}}^{k_{\rm max}}\cos\left(M\phi(\tau)\right)\,\Re\hat{n}\left(\gamma+\frac{iM}{t}\phi(\tau)\right)\phi'(\tau),
\label{numerics1}
\ee
where $k_{\rm max}$ is an integer and $\tau=kh+\frac{\pi}{2M}$. We note that $\phi'(\tau)\to0$ double exponentially as $\tau\to-\infty$. We see that $\phi(\tau)\to\tau$ double exponentially as $\tau\to\infty$ and 
$\cos(M\phi(kh+\frac{\pi}{2M}))\sim\cos(Mkh+\frac{\pi}{2})=0$.

\begin{figure}[ht]
\begin{center}
\includegraphics[width=0.7\textwidth]{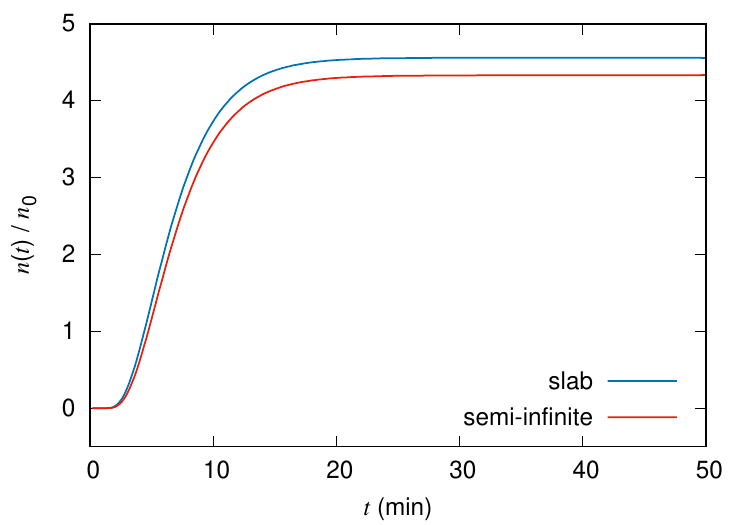}
\end{center}
\caption{
The particle number density is plotted as a function of $t$. The blue curve is from (\ref{numerics1}) and the red curve is from (\ref{numerics2}).
}
\label{fig1}
\end{figure}

We set $L=10\,{\rm cm}$, $N=30$, $k_{\rm max}=50$, and $M=50$. We found $\gamma=0.04$ is suitable. Time was discretized as $t_j=j\Delta t$ ($\Delta t=0.2\,{\rm min}$, $j=1,\dots,250$). Furthermore we set $\sigma_a=10^{-8}\,{\rm min}^{-1}$, $\sigma_s=5\,{\rm min}^{-1}$, $v_0=5\,{\rm cm}/{\rm min}$, and $u=1.5\,{\rm cm}/{\rm min}$. The computation time was $90\,{\rm sec}$ on a laptop computer (MacBook Pro, 2.3 GHz Intel Core i5). The result is plotted in Fig.~\ref{fig1}. In \cite{Amagai-etal20}, the particle number density was defined by
\be
n(x,t)=\int_{-1}^1\psi_{\infty}(x,\mu,t)\,d\mu,
\label{numerics2}
\ee
where $\psi_{\infty}(x,\mu,t)$ is the solution to (\ref{rte0}) when $L\to\infty$. Figure \ref{fig1} also shows $n(L,t)/n_0$ for comparison.

\begin{rmk}
According to the error analysis in \cite{Ganapol08}, $\gamma t$ must be small. In \cite{Ganapol08}, it is suggested to move $\gamma$ according to $t$, and the form $\gamma=\bar{\gamma}+\alpha/t$ is proposed, where $\bar{\gamma},\alpha$ are positive constants.
\end{rmk}

\section{Concluding remarks}
\label{concl}

Although the solution of the transport equation well described the experimentally obtained breakthrough curve in \cite{Amagai-etal20}, the half space was assumed in the formulation. To take into account the length of the column, we gave a formulation in the slab geometry, which has both ends and tracer particles enter from one end and exit from the other end. For the numerical inversion of the Laplace transform, we could apply the double-exponential formula after expressing $n(t)$ using $\cos\omega$ in (\ref{ntcosw}).

Since sands and glass beads are packed with an equal density in the column, isotropic scattering seems to be a reasonable assumption. However, there is no reason to exclude the possibility of anisotropy. The introduction of the anisotropy factor ${\rm g}\neq0$ is a future issue.

The most precise geometry for the column experiment is a cylinder in three dimensions. However, the essential nature of the transport is expected to be seen by the one-dimensional transport equation since the experimental setup is designed so that the flow is identical in horizontal directions.

\section*{Acknowledgements}
MM acknowledges support from Grant-in-Aid for Scientific Research (17K05572, 18K03438) of JSPS.

%\newpage                                                            
%\setcounter{section}{1}
%\appendix

%\section*{References}

\end{document}